\documentclass[sigplan]{acmart}

\makeatletter
\def\@ACM@checkaffil{
    \if@ACM@instpresent\else
    \ClassWarningNoLine{\@classname}{No institution present for an affiliation}%
    \fi
    \if@ACM@citypresent\else
    \ClassWarningNoLine{\@classname}{No city present for an affiliation}%
    \fi
    \if@ACM@countrypresent\else
        \ClassWarningNoLine{\@classname}{No country present for an affiliation}%
    \fi
}
\makeatother

\renewcommand\footnotetextcopyrightpermission[1]{}
\usepackage{tikz}
\usepackage{amsmath}
\usepackage{caption}
\usepackage{capt-of}
\usepackage{cleveref}
\crefname{section}{§}{§§}
\Crefname{section}{Section}{Sections}
\usepackage[subtle]{savetrees}
\usepackage{placeins}
\usepackage{tabularx}

\renewcommand\_{\textunderscore\allowbreak}

\begin{document}
\sloppy

\title{Unikernel Linux (UKL)}

\author{Ali Raza}
\affiliation{\institution{Boston University}}
\email{aliraza@bu.edu}

\author{Thomas Unger}
\affiliation{\institution{Boston University}}
\email{tommyu@bu.edu}

\author{Matthew Boyd}
\affiliation{\institution{MIT CSAIL}}
\email{mcboyd@mit.edu}

\author{Eric B Munson}
\affiliation{\institution{Boston University}}
\email{munsoner@bu.edu}

\author{Parul Sohal}
\affiliation{\institution{Boston University}}
\email{psohal@bu.edu}

\author{Ulrich Drepper}
\affiliation{\institution{Red Hat}}
\email{drepper@redhat.com}

\author{Richard Jones}
\affiliation{\institution{Red Hat}}
\email{rjones@redhat.com}

\author{Daniel Bristot de Oliveira}
\affiliation{\institution{Red Hat}}
\email{bristot@redhat.com}

\author{Larry Woodman}
\affiliation{\institution{Red Hat}}
\email{lwoodman@redhat.com}

\author{Renato Mancuso}
\affiliation{\institution{Boston University}}
\email{rmancuso@bu.edu}

\author{Jonathan Appavoo}
\affiliation{\institution{Boston University}}
\email{jappavoo@bu.edu}

\author{Orran Krieger}
\affiliation{\institution{Boston University}}
\email{okrieg@bu.edu}

\renewcommand{\shortauthors}{Raza et al.}

\begin{abstract} 
This paper presents Unikernel Linux (UKL), a path toward integrating unikernel
optimization techniques in Linux, a general purpose operating system.  UKL adds
a configuration option to Linux allowing for a single, optimized process to link
with the kernel directly, and run at supervisor privilege. This UKL process does
not require application source code modification, only a re-link with our,
slightly modified, Linux kernel and  \texttt{glibc}. Unmodified applications
show modest performance gains out of the box, and developers can further
optimize applications for more significant gains (e.g. 26\% throughput
improvement for Redis).  UKL retains support for co-running multiple user level
processes capable of communicating with the UKL process using standard IPC.  UKL
preserves Linux's battle-tested codebase, community, and ecosystem of tools,
applications, and hardware support.  UKL runs both on bare-metal and virtual
servers and supports multi-core execution. The changes to the Linux kernel are
modest (1250 LOC).  
\end{abstract}

\maketitle


\section{Introduction} \label{sec:intro}

There is growing evidence that the structure of today's general-purpose
operating systems is problematic for a number of key use cases. For example,
applications that require high-performance I/O often bypass the kernel by using
frameworks like DPDK \cite{dpdk} and SPDK \cite{spdk} to gain direct access
to hardware devices~\cite{intel, netmap}. In the cloud, there is a mismatch
between client workloads and the kernels that support them: workloads
are often single-user and single-process, while general purpose OSes
support multi-user and multi-process environments~\cite{serverlesspeeking}.

In response, there has been a resurgence of research systems exploring the idea
of a libraryOS, or \textit{a unikernel}, where an application is linked with a
specialized kernel and deployed directly on virtual hardware~\cite{exokernel}.
Compared with Linux, unikernels have demonstrated significant advantages in boot
time~\cite{mirageOS,serverlessenddominance}, security~\cite{nabla}, resource
utilization~\cite{lightVM,seuss}, and I/O performance~\cite{ebbrt}.

As with any operating system, widespread adoption of a unikernel will require
significant and ongoing investment by a large community. Justifying this
investment is difficult because unikernels target only niche portions of the
broad use cases of general-purpose OSes. In addition to their intrinsic
limitation as single application environments, with few exceptions, existing
unikernels support only virtualized environments. Further, in many cases, they
only run on a single processor core. Finally, they do not support accelerators
(e.g., GPUs and FPGAs) that are increasingly critical to achieving high
performance in a post-Dennard-scaling world.

Some systems have demonstrated that it is possible to create a unikernel that
reuses much of the battle-tested code of a general-purpose OS and supports a
wide range of applications. Examples include NetBSD-based Rump Kernel
\cite{rumpkernel}, Windows-based Drawbridge \cite{drawbridge}, and Linux-based
Linux Kernel Library (LKL)~\cite{lkl}. These systems, however, require
significant changes to the general-purpose OS, resulting in a fork of the
codebase and community. As a result, ongoing investments in the base operating
system are typically not incorporated into the forked unikernels.

To avoid the investment required to create a new OS, the recent
Lupine~\cite{lupine} and X-Containers~\cite{xcontainers} projects exploit
Linux's innate configurability to enact application-specific customization.
These projects avoid the hardware overhead of system calls between user and
kernel mode but do not explore deeper optimizations. Essentially these systems
preserve the API between the application and the underlying kernel, giving up on
unikernel performance advantages that depend on linking the application and
kernel code together.

The Unikernel Linux (UKL) project started as an effort to exploit Linux's
configurability to try to create a new unikernel in a fashion that would avoid
forking the kernel. If this were possible, we hypothesized that we could create
a unikernel that would support a wide range of Linux applications and hardware
while becoming a standard part of the ongoing investment by the Linux community.
Our experience has led us to a more general goal: creating a kernel that can be
configured to span the spectrum between a general-purpose operating system, amenable to
a large class of applications, and a highly optimized, possibly application and
hardware specialized unikernel. 

If all UKL configurations are disabled, a standard Linux kernel is generated.
When the \textit{base model} UKL configuration is used, we start on the general
purpose end of the spectrum. This simplest configuration of UKL supports many
applications, albeit with only modest performance advantages. Like many
unikernels, a single application is statically linked with the kernel and
executed in supervisor mode. However, the base model of UKL preserves most of
the invariants and design of Linux, including a separate page-able application
portion of the address space and a pinned kernel portion, distinct execution
modes for application and kernel code, and the ability to run multiple
processes.   As a result, this base model provides an avenue toward supporting
all hardware and applications of the original kernel and the entire Linux
ecosystem of tools for deployment, debugging, and performance tuning. Both the
changes to Linux to support the UKL base model (\textasciitilde550
LOC\footnote{UKL base model was submitted as a `Request for Comments' (RFC) to
the Linux community}) and the resulting performance improvement (e.g., 5\% for
syscall) are modest. 

Once an application runs in the UKL base model, a developer can move along the
spectrum towards a specialized unikernel by 1)~adapting additional configuration
options that may improve performance but will not work for all applications
and/or 2)~modifying the applications to directly invoke kernel functionality.
Example configuration options we have explored avoid costly transition checks
between application and kernel code, use \texttt{ret} instruction (rather than
\texttt{iret}) to return from page faults and interrupts, and use shared stacks for
application and kernel execution. Application modifications can, for example,
avoid scheduling and exploit application knowledge to reduce the overhead of
synchronization and polymorphism. Experiments show up to 83\% improvement in
syscall latency and substantial performance advantages for real workloads, e.g.,
26\% improvement in Redis throughput while improving tail latency by 22\%. The
full UKL patch to Linux, including the base model and all configurations, is
1250 LOC. 
 
Our research focuses on runtime performance and not the many other benefits
(e.g., security, boot time, resource utilization) demonstrated by other
unikernel research. With that caveat, contributions of this work include:

\begin{enumerate} 

\item An existence proof that unikernel techniques can be integrated into a
general-purpose OS in a fashion that does not need to fragment/fork it.  

\item A demonstration that a single kernel can be adopted across a spectrum
between a general-purpose OS and a specialized unikernel.

\item A demonstration that performance advantages are possible; applications
achieve modest gains with no changes, and incremental effort can achieve more
significant gains.  

\end{enumerate}
 


\section{Motivation \& Goals} \label{sec:goals} 

UKL seeks to explore a spectrum of unikernel optimization levels. At one end is
a general-purpose operating system supporting a wide class of applications and
hardware. At the other is a highly optimized unikernel which may be specialized
down to the specifics of a single application and platform. In doing this, we
aim to (1) enable unikernel optimizations demonstrated by earlier systems while
preserving a general-purpose operating system's (2) broad application support,
(3) broad hardware support, and (4) the ecosystem of developers, tools, and
operators. We motivate and describe each of these four goals.

\subsection{Unikernel optimizations} \label{sub:goal-uklmod}

Unikernels fundamentally enable optimizations that rely on linking the
application and kernel together in the same address space. Example optimizations
that previous systems have adopted include 1)~avoiding ring transition overheads
\cite{lupine,kml}; 2)~exploiting the shared address space to pass pointers
rather than copying data \cite{ebbrt}; 3)~exploiting fine-grained control over
scheduling decisions, e.g., deferring preemption in latency-sensitive routines;
4)~enabling interrupts to be efficiently dispatched to application code
\cite{ebbrt}; 5)~exploiting knowledge of the application to remove code that is
never used \cite{mirageOS}; 6)~employing kernel-level mechanisms to optimize
locking and memory management \cite{osv}, for instance, by using
Read-Copy-Update (RCU)~\cite{rcu}, per-processor memory, and DMA-aided data
movement; and 7)~enabling compiler, link-time, and profile-driven optimizations
between the application and kernel code.

Ultimately our goal with UKL is to enable the full spectrum between
general-purpose and highly specialized unikernels. For this paper, our goal is
to enable applications to be linked into the Linux kernel and explore what
performance improvements can be achieved, first in the context of unmodified
application source plus a re-compilation and link step, then by modest changes
to the application and general-purpose system.

\subsection{Application support} \label{sec:goal_app} 

One of the fundamental problems with unikernels is the limited set of
applications that they support. In their purest form, unikernels only execute a
single process; this excludes any application that requires helper processes,
scripts, etc. Moreover, the limited set of interfaces typically requires
substantial porting effort for any application and libraries that the
application uses.

UKL seeks to enable unikernel optimizations while remaining broadly compatible.
Our goal is to enable any unmodified Linux application and library to use UKL
with a re-compilation, as long as only one application needs to be linked into
the kernel. Once the application runs as a unikernel, the developer can
incrementally enable unikernel optimizations/configurations. A large set of
applications should be able to achieve some gain on the general-purpose end of
the spectrum, while a much smaller set of applications will be able to achieve
more substantial gains as we move toward the specialized unikernel end.

\subsection{Hardware support} \label{sec:goal_hw} 

Another fundamental problem with unikernels is the lack of support for physical
machines and devices. While recent unikernel research has mostly focused on
virtual systems, some recent~\cite{ebbrt,rumprun} and
previous~\cite{bluegene,cachekernel,exokernel,nemesis,andersoncase} systems have
demonstrated the value of per-application specialized operating systems on
physical machines. Unfortunately, even these systems were limited to very
specific hardware platforms with a restricted set of device drivers. This
precludes a wide range of infrastructure applications (e.g., storage systems,
schedulers, networking toolkits) that are typically deployed bare metal.
Moreover, the lack of hardware support is an increasing problem in a
post-Dennard scaling world, where performance depends on taking advantage of the
revolution of heterogeneous computing.

Our goal with UKL is to provide a unikernel environment capable of supporting
the complete hardware compatibility list of Linux, allowing applications to
exploit any hardware (e.g., GPUs, TPUs, FPGAs) enabled in Linux. Our near-term
goal, while supporting all Linux devices, is to focus on x86-64 systems. Much
like KVM became a feature of Linux on x86 and was then ported to other
platforms, we expect that, if UKL is accepted upstream, communities interested
in non-x86 architectures will take on porting and optimizing UKL for their
platforms.

\subsection{Ecosystem} \label{sec:goal_eco} 

While application and hardware support are normally considered the fundamental
barriers to unikernel adoption, the problem is much larger. Linux has a huge
developer community, operators that know how to configure and administer it, a
massive body of battle-tested code, and a rich set of tools to support
functional and performance debugging and configuration.

Our goal with UKL is to preserve the entire ecosystem on the general-purpose end
of the spectrum while enabling developers to adopt extreme optimizations
inconsistent with the broader ecosystem. This means operational, functional, and
performance debugging tools should just work. Standard application and library
testing systems should, similarly, just work.\footnote{In fact, we used a great
deal of \texttt{glibc} and libpthread's internal unit tests to identify and fix
problems within UKL.} Most of all, the base changes to enable UKL must preserve
the assumptions of the battle-tested Linux code, must be testable and
maintainable as development on the system progresses, and must be in a form that
can be accepted by the community.


\section{Design} \label{sec:design}

UKL's {\tt base model} preserves almost all the properties of the
general-purpose operating systems; importantly, the base model preserves the
(known and unknown) invariants and assumptions of applications and Linux, except
system calls are replaced by function calls and application code is linked with
kernel code and executes in kernel mode.  From this starting point, an expert
programmer can adopt specific unikernel optimizations that are valuable and safe
for their specific application by choosing (additional) configuration options
and/or modifying the application to invoke kernel functionality directly. Here
we describe the base model and then some of the unikernel optimizations we have
explored.

\subsection{Base Model} 

UKL is similar to many unikernels in that it involves modifications to a base
library and a kernel and has a build process that enables a single application
to be statically linked with kernel code to create a bootable kernel. In the
case of UKL, the modifications are to \texttt{glibc} and the Linux kernel. As a
result of the wide variety of architectures supported by \texttt{glibc} and
Linux, it was possible to introduce the majority of changes we required in a new
UKL target architecture; most of the hooks we require to override code already
exist in the common code of these projects.

The base model of UKL differs from unikernels in 1) support for multiple
processes, 2) address space layout, and 3) maintaining distinct execution models
for applications and the kernel.

\paragraph{Support for multiple processes:} One key area where UKL differs from
unikernels is that, while only one application can be linked into the kernel,
UKL enables other applications to run unmodified on top of the kernel. Support
for multiple processes is critical to support many applications that are
logically composed of multiple processes (\cref{sec:goal_app}), standard
configuration and initialization scripts for device bring-up
(\cref{sec:goal_hw}), and the tooling used for operations, debugging, and
testing (\cref{sec:goal_eco}).

It is important to note that while UKL supports multiple processes,  other
processes are not protected from the performance-optimized one linked into the
kernel. Similar to unikernels, our security model assumes other work that needs
to be protected is isolated by a hypervisor or by techniques for securing bare
metal machines~\cite{bolted,nitro}.

\begin{figure}[]
    \includegraphics[scale=0.18]{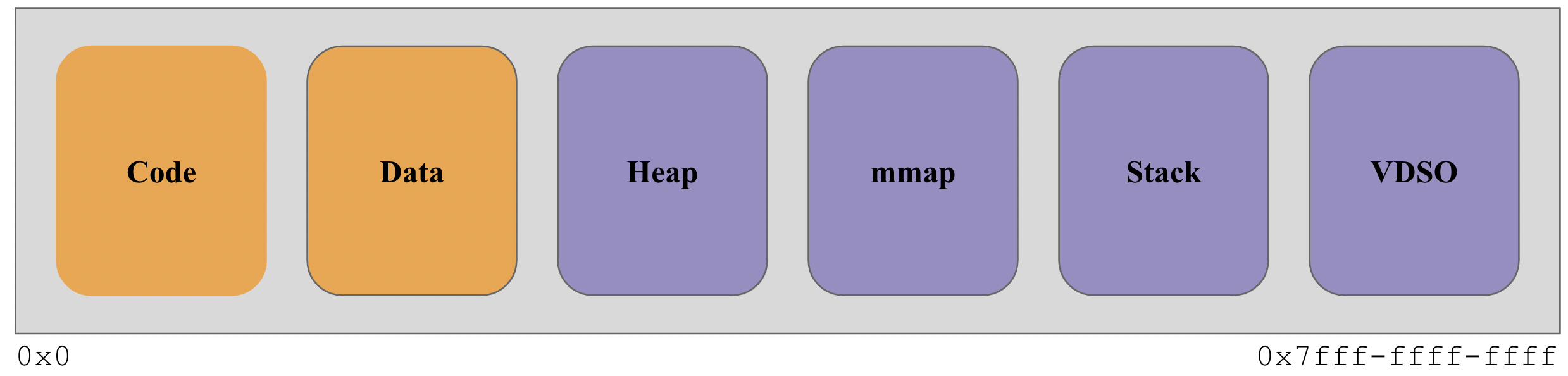}
    \caption{The 64-bit virtual address space layout for non-UKL user process.
    In orange are the segments that are relocated to the kernel half for the UKL
    process.
    } 
    \vspace{-0.2in}
    \label{fig:user_vas}
\end{figure}

\paragraph{Address space layout:} UKL preserves the standard Linux virtual
address space split between application and kernel. The application heap,
stacks, and mmapped memory regions are all created in the user portion of the
address space. Kernel data structures (e.g., task structs, file tables, buffer
cache) and kernel memory management services (e.g., vmalloc and kmalloc) all use
the kernel portion of the address space. Since the kernel and application are
compiled and linked together, the application (and kernel) code and data are all
allocated in the kernel portion of the virtual address space. See figure \ref{fig:user_vas}.

We found it necessary to adapt this address space layout because Linux performs
a check to see if an address being accessed is pinned or not; modifying this
layout would have resulted in changes that may have been difficult to upstream
(\cref{sec:goal_eco}). Unfortunately, this layout has two negative implications
for application compatibility. First, (see~\cref{sec:imp}) applications must be
compiled with different flags to use the higher portion of the address space.
Second, it may be problematic for applications with large initialized data
sections that, in UKL, are now pinned.

\paragraph{Execution models:} \label{par:exec_mod} Even though the application
and kernel are linked together, UKL differs from unikernels in providing
fundamentally different execution models for application and kernel code.
Application code uses large stacks (allocated from the application portion of
the address space), is fully preemptable, and uses application-specific
libraries. This model is critical to supporting a large set of applications
without source modification (\cref{sec:goal_app}).

On the other hand, kernel code runs on pinned stacks, accesses pinned data
structures, and uses kernel implementation of common routines. This model was
required to avoid substantial modifications to Linux that may prohibit
acceptance by the community (\cref{sec:goal_eco}).

\begin{figure}[]
    \includegraphics[scale=0.25]{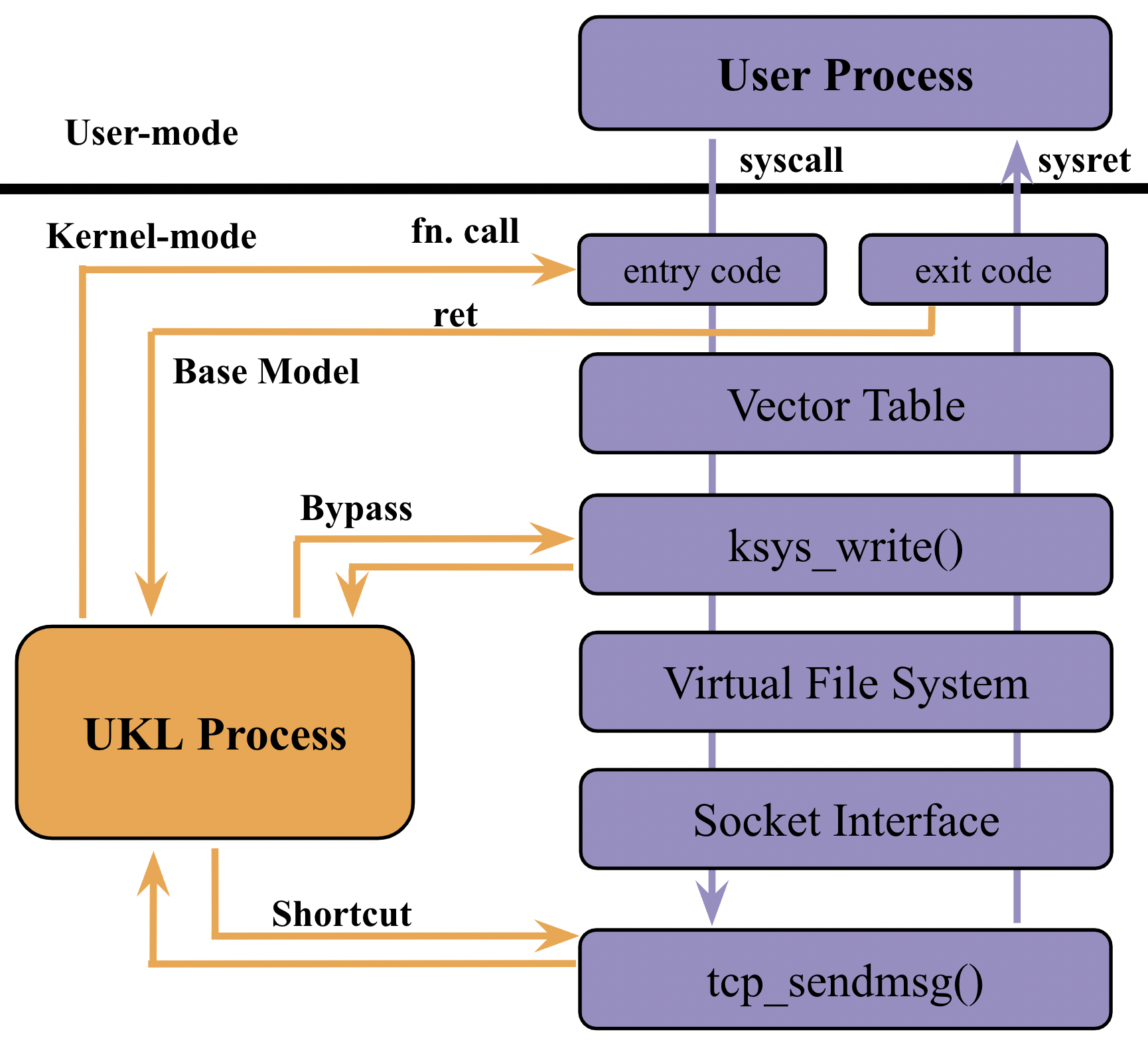}
    \caption{
       A schematic of a \texttt{write} system call destined for a network device.
       The three alternative internal entry points that a UKL process exercises
	are shown in orange.
    } 
    \vspace{-0.2in}
    \label{fig:entries}
\end{figure}

On the transition between the execution models, UKL performs the same entry and
exit code of the Linux kernel, with the difference that: 1)~transitions to
kernel code are done with a procedure call rather than a \texttt{syscall}{}, and
2)~transitions from the kernel to application code are done via a \texttt{ret}
rather than a \texttt{sysret}{} or \texttt{iret}; see "Base Model" figure \ref{fig:entries} . This transition code includes
changing between application and kernel stacks, RCU handling, checking if the
scheduler needs to be invoked, and checking for signals. In addition, it
includes setting a per-thread \texttt{ukl\_mode} to identify the current mode of
the thread so that subsequent interrupts, faults, and exceptions will go through
normal transition code when resuming interrupted application code. This is how
we minimize affecting kernel invariants in the base model.

\subsection{Unikernel Optimizations}

While preserving existing execution modes enables most applications to run with
no source modifications on UKL, the performance advantages of just avoiding
\texttt{syscall}{}, \texttt{sysret}{}, and \texttt{iret} operations are, as
expected, modest. However, once an application is linked into the kernel,
different unikernel optimizations are possible. First, a developer can apply a
number of configuration options that may improve performance. Second, a
knowledgeable developer\footnote{Expertise is needed to perform these
customizations. For example, the kernel will fail if an application calls an
internal kernel routine, passing a pointer to an application data structure that
resides on a page that has not yet been accessed/allocated. We are just starting
to develop a body of use cases and examples that should inform developers about
the care they should take for different optimizations.} can improve performance
by modifying the application to call internal kernel routines and violating, in
a controlled fashion, the normal assumptions and invariants of kernel versus
application code.

\subsubsection{Configuration Options} 

Once an application is running, a developer can easily explore a number of
configuration options that, while not safe for {\bf all} applications, may be
safe and offer performance advantages for their application.  

\paragraph{Bypassing entry/exit code:} On Linux, whenever control transitions
between application and kernel through system calls, interrupts, and exceptions,
some entry and exit code (\cref{par:exec_mod} Execution models) is executed; this
is expensive. We introduced a configuration (UKL\_BYP) that allows the
application, on a per-thread basis, to tell UKL to bypass entry and exit code
for some transitions between application and kernel code; see "Bypass" figure \ref{fig:entries}. As we will see, this
model results in significant performance gains for applications that make many
small kernel requests.

A developer can invoke an internal kernel routine directly, where no automatic
transition paths exist, e.g., invoking \texttt{vmalloc} to allocate pinned
pre-allocated kernel memory rather than normal application routines. Using such
memory avoids subsequent faults and results in less overhead when kernel
interfaces have to copy data to and from that memory. 

\paragraph{Avoiding stack switches:} Linux runs applications on dynamically
sized user stacks and kernel code on fixed-sized, pinned kernel stacks. Every
time kernel functionality is invoked, this stack switch breaks the compiler's
view and limits cross-layer optimizations, e.g., link-time optimizations,
etc.\footnote{Unfortunately, today LTO in Linux is only possible with CLANG
while \texttt{glibc} can only be compiled with GCC. Efforts are underway in the
community to enable \texttt{glibc} to be compiled with CLANG and enable Linux
LTO with GCC. We are excited to explore the advantages of LTO as soon as one of
these efforts completes.} The developer can select between two UKL
configurations that avoid stack switching(UKL\_NSS and UKL\_NSS\_PS), where (see
implementation) each is appropriate for a different class of application. 

\paragraph{\texttt{ret} versus \texttt{iret}:} Linux uses \texttt{iret} when
returning from interrupts, faults and exceptions. \texttt{iret} is an expensive
instruction that atomically changes the privilege level, instruction pointer,
stack pointer, etc. UKL\_RET configuration option uses \texttt{ret} and ensures
atomicity by enabling interrupts only after returning to the application stack. 

\subsubsection{Application Modifications} 

Along with the above configurations, developers can explore deeper optimizations
by taking advantage of application knowledge. For example, they may be able to
assert that only one thread is accessing a file descriptor and avoid costly
locking operations. As another example, they may know \textit{a priori} that an
application uses TCP and not UDP and that a particular write operation in the
application will always be to a TCP socket, avoiding the substantial overhead of
polymorphism in the kernel's VFS implementation. As we optimize specific
operations, we are building up a library of helper functions that simplify
common operations; see "Shortcut" figure \ref{fig:entries}.

UKL base model ensures that the application and kernel execution models stay
separate, with proper transitions between the two. Applications can toggle a
per-thread flag which switches them to the kernel-mode execution, allowing
application threads to be treated as kernel threads, so they won't be preempted.
This can be used as a `run-to-completion' mode where performance-critical paths
of the application can avoid perturbation.
\vspace{-0.15in}

\section{Implementation} \label{sec:imp} 

The size of the UKL base model patch to Linux kernel 5.14 is approximately 550
lines, and the full UKL patch (base model plus all the configuration options) is
1250 lines. Most of these changes are target-specific, i.e., in the x86
architecture directory.

UKL takes advantage of the existing kernel Kconfig and \texttt{glibc} build
systems. These allow target-specific functionality to be introduced that doesn't
affect generic code or code for other targets. All UKL changes are wrapped in
macros which can be turned on or off through kernel and \texttt{glibc} build
time config options; they are compiled out when Linux and \texttt{glibc} are
configured for a different target. 

We found that the UKL patch can be small due to recent favorable design decisions
by the Linux community. For instance, Linux's low-level transition code has
recently undergone massive rewriting to reduce assembly code and move
functionality to C language. This has allowed UKL transition code changes to be
localized to that assembly code. Further, the ABI for application threads
dedicates a register (\texttt{fs}) to point to thread-local storage, while
kernel threads have no such concept but instead dedicate a register
(\texttt{gs}) to point to processor-specific memory. If a register was used by
both Linux and \texttt{glibc}, UKL would have had to add code to save and
restore it on transitions; instead, both registers can be preserved.

In addition to the kernel changes, about 5,439 lines of code are added or
changed in \texttt{glibc}. This number is inflated because according to the
\texttt{glibc} development approach, any file that needs to be modified has to
be first copied to a new sub-directory and then modified. The actual number of
lines changed in \texttt{glibc} is 1,720. All the UKL changes are well contained
in a separate directory. The \texttt{glibc} build process, configured for UKL,
first searches the UKL specific directory for a target file at build time before
searching the default location.

\paragraph{Building UKL binary:} Our current implementation does not support
dynamically loaded libraries, requiring the application code and associated user
libraries to be compiled and statically linked with the kernel. This code is
built with two special flags. The first flag disables the red zone optimization
(\texttt{-mno-red-zone}) as is standard when building kernel code. The second
kernel memory model flag (\texttt{-mcmodel=kernel}) enables the generated code
to be loaded into the highest 2GB of address space instead of the lower 2GB, the
default for user code.

While a limitation of our current implementation that attempts to minimize
changes to Linux, we do believe future work could enable binary compatibility.
The kernel already supports dynamic linking, and extending the dynamic linker to
support application libraries seems feasible. Previous work~\cite{hermitux} has
shown how all faults, interrupts, and exceptions can be made to use dedicated
stacks through the Intel interrupt stack table (IST); such an approach would
eliminate the need to disable red zones. Finally, the application and libraries
could be loaded into the lower 2GB of address space, with some extra work at the
transition between application and kernel code. 

After the application and libraries are compiled, a modified kernel build system
combines them and the kernel into a final \texttt{vmlinux} binary which can be
booted bare-metal or virtual. To avoid name collisions, all application symbols
(including library ones) are prefixed with \texttt{ukl\_} using the
\texttt{objcopy} utility before linking the application and kernel together.
Kernel code typically has no notion of thread-local storage or C++ constructors,
so the kernel's linker script is modified to link with userspace code and ensure
that thread-local storage and C and C++ constructors and destructors work.
Appropriate changes to the kernel loader are also made to load the new ELF
sections along with the kernel. 

\paragraph{Transition between application and kernel code:} On transitions
between application and kernel code, the normal entry and exit code of the Linux
kernel is executed, with the main change being that transitions code use
\texttt{call/ret} instead of \texttt{syscall/sysret}. The \texttt{syscall}
instruction puts return address in \texttt{rcx} register, but \texttt{call} puts
it on the user stack. After the register state is pushed to the kernel stack in
UKL, the return address on the kernel stack needs to be updated with the correct
address.

The different configurations of UKL involve changes to the transitions between
application and kernel code. All changes were made through Linux
\texttt{SYSCALL\_DEFINE} and \texttt{glibc} \texttt{INLINE\_SYSCALL} macros. For
example, to enable UKL\_BYP mode, we add a stub in the kernel macro that is
invoked by the corresponding \texttt{glibc} macro. A per-thread flag
(\texttt{ukl\_byp}) is used to identify if the bypass optimization is turned on
or off for that thread.

Linux tracks whether a process runs in user or kernel mode through the
\texttt{CS} register, but UKL-optimized applications are always in kernel mode.
So the UKL thread tracks this in a flag (\texttt{ukl\_mode}) in the kernel's
thread \verb|task_struct|. 

The UKL\_RET configuration option replaces \texttt{iret} after the application
code is interrupted by a page fault or an interrupt, with a \texttt{ret}. A
challenge is that we cannot enable interrupts until we have switched from the
kernel stack to the application stack, or the system might land in an undefined
state. To support this, UKL copies the return address and user flags from the
kernel stack to the user stack, switches to the user stack, and only then pops
the flags (which might enable interrupts), and does a \texttt{ret} to the
application code.

\paragraph{Enabling shared stacks:} In the UKL base model, a switch between the
user and kernel stack occurs when code transitions between the two domains,
limiting cross-layer compiler optimization. We have developed two configurations
to avoid these stack switches. 

The UKL\_NSS configuration option simply executes kernel code on the application
stack. With this configuration, while applications can run before and after the
UKL application, the developer must ensure that other applications do not run
concurrently with the UKL application. In the case of an inter-processor
interrupt, e.g., for a TLB invalidation, Linux can store information on the
current process stack. If the interrupt interrupts a non-UKL thread, the kernel
will inherit that other process's page tables and then try to access the
information stored on the UKL thread's user stack that is not mapped in the
current page tables, resulting in a kernel panic. 

The UKL\_NSS\_PS configuration option allocates fixed-sized stacks in the kernel
part of the address range. This configuration allows multiple processes to run
concurrently but is impractical for applications that assume many threads with
huge virtual stacks, pre-allocating and pinning all that memory in the kernel
address space.

\paragraph{Page-faults:} If the UKL\_NSS configuration option is on, if a page
fault occurs, it can result in two problems. First, deadlocks can occur if the
fault happens when kernel memory management code is being executed, e.g.,
\texttt{mmap}. If the thread is holding a lock on the memory control struct
(\texttt{mm\_struct}), then the page fault handler will deadlock when it tries
to acquire that same lock to read which virtual memory area (VMA) the faulting
address belongs to. To address this problem, UKL saves a reference to the user
stack VMA when a UKL thread or process is created. In case of page faults, while
user stacks are used throughout, we first check if the faulting address is a
stack address by comparing it against the address range of a saved VMA. If so,
we know it's a stack address, and the code knows how to handle it without taking
any further lock. If not, we first take a lock to retrieve the correct VMA and
move forward normally.

Second, since in kernel mode the hardware simply saves the state on the current
(user) stack, the kernel will fail on a double fault. The UKL\_PF\_DF
configuration option addresses this by adding code to the double fault handler
to check if the fault is to a stack page, and if so, branch to the regular page
fault code. Alternatively, the UKL\_PF\_SS configuration option solves this
problem by updating the IDT to ensure that the page fault handler always
switches to a dedicated stack through the Interrupt Stack Table (IST) mechanism. 

\paragraph{Fork, Clone:} Userspace processes running on UKL can \texttt{fork} as
expected. We have implemented a modified \texttt{fork} for the UKL-optimized application,
where any state in the user portion of the address space is treated as expected,
but the data in the kernel portion of the address space is shared. We have used
this, for example, to run multiple optimized UKL processes for the LEBench
microbenchmark~\cite{lebench}. Again, loading process data into the user portion
of the address space would enable full \texttt{fork} support. 

To create UKL threads, the userspace pthread library runs
\texttt{pthread\_create}, which calls \texttt{clone}. We modified this library
to pass a new flag \texttt{CLONE\_UKL} to ensure the correct initial register
state is copied into the new task either from the user stack or kernel stack,
depending on whether the parent is configured to switch to the kernel stack or
not. 

\paragraph{Changes to \texttt{execve}:} Userspace processes running on UKL can
\texttt{exec} as expected.  The UKL process is started by invoking \texttt{exec} with a
program name specified by a configuration; again, if called twice, it will
result in two processes sharing the portion of data in the kernel address space.
While most of \texttt{execve} is unmodified, we skip loading the (nonexistent)
binary and jump directly to the \texttt{glibc} entry point.  \texttt{glibc}
initialization happens almost as normal, but when initializing thread-local
storage, changes had to be made to read symbols set by the kernel linker script
instead of an ELF binary. C and C++ constructors run the same way as in a normal
process. Command-line parameters to \verb|main| are extracted from a part of the
Linux kernel command line. While we have not yet done so, we believe only modest
effort is required to enable a UKL-optimized process to call \texttt{fork} and
then \texttt{exec} a (non-optimized) application. 


\section{Evaluation} \label{sec:evaluation}

The main goals of UKL are to integrate optimizations explored by different
unikernels into a general-purpose operating system while preserving its
application and hardware compatibility and its ecosystem of tools, utilities,
and community of developers. We want to evaluate if UKL can
achieve those goals and if there are any performance advantages compared to
Linux and other unikernels or library operating systems.

We discuss our experience with UKL preserving Linux's application and hardware
compatibility, and
ecosystem in \cref{eval:lnx_compat}. In \cref{eval:Micro},
microbenchmarks are used to evaluate the performance of UKL on simple system
calls (\cref{eval:sc-base}), more complex system calls (\cref{eval:sc-large})
and page faults (\cref{eval:PFH}). We also explore if any improvements in the
microbenchmarks translate to speed-ups in I/O (\cref{eval:fio}) and find a 36\%
improvement in a latency-sensitive benchmark. We find that the advantage of
adopting unikernel optimizations is large for simple kernel calls (e.g., 83\%)
and still significant for expensive kernel calls that transfer 8KB of data
(e.g., 24\%).  

\Cref{eval:redis,eval:memcached} evaluate UKL performance for a single-threaded
application virtualized and bare metal as well as a complex multi-threaded application
in a virtualized setting. We find that UKL can enable significant throughput (e.g., up to
26\%) and latency (e.g., up to 22\%) advantages over Linux.

\subsection{Linux application hardware \& ecosystem} \label{eval:lnx_compat} 

The fundamental goals of the UKL project are to integrate unikernel
optimizations without losing Linux's broad support for applications, hardware,
and ecosystem.

\paragraph{Application support:} 

After compilation and linking, we expected no significant challenges in running
different unmodified applications as optimization targets with the UKL base
model. Our hypothesis was largely true. We tested dozens of unmodified
applications without any additional UKL-specific effort, including Memcached
\cite{memcached}, Redis \cite{redis}, Nginx \cite{nginx}, FIO \cite{fio}, a
multi-party computation benchmark \cite{secrecy}, a small TCP echo server,
simple programs to test C++ constructors and the C++ Standard Template Library
(STL), the GAP Benchmark Suite~\cite{DBLP:journals/corr/BeamerAP15} (a complex
C++ graph based benchmark suite) and LEBench \cite{lebench} (a Linux system call
benchmark).

Although the experience of running different applications with UKL was largely
smooth, we did experience three (anticipated) challenges. First, it can be
challenging to recompile and statically link applications with complex
dependencies and Makefiles. Second, we have hit some programs that by default
invoke \texttt{fork} followed by \texttt{exec}, e.g., Postgres, or depend on the
dynamic loader. Third, we have run into issues with proprietary applications
available in only binary form, e.g., user-level libraries for GPUs. 

\paragraph{Hardware support:} For hardware, we have not run into any
compatibility issues and have booted or \texttt{kexec}-ed to UKL on a wide variety of
x86-64 servers, virtualization platforms, and laptops with different Intel,
Brocade and virtual NICs, as well as NVMe, SATA controllers, and virtual block
devices. The scripts and tools used to deploy and manage normal Linux machines
were used for UKL deployments without modification. Although we could not
exploit GPUs without access to the source code for key libraries, we enabled UKL
on Linux and deployed it bare-metal, which is the first step in enabling
accelerators. We expect no major issues using accelerators with UKL if we have
access to source code or binaries compiled with UKL-specific flags (see
\cref{sec:imp}).

\paragraph{Ecosystem:}  We have been able to run all the different applications,
utilities, and tools that can run on unmodified Linux. This has been extremely
critical in building UKL, i.e., we use all the debugging tools and techniques
available in Linux. We have been able to profile UKL workloads with
\texttt{perf} and identify code paths that could be squashed for
performance benefits (see \cref{fig:flamegraph}).

The UKL patch size for the base model is around 550 LOC, and the full UKL patch
with all the configurations we have explored so far is 1250 lines. We have spent
several months discussing and presenting the concept of unikernels and the UKL
approach to kernel developers within Red Hat. We posted the base model as an RFC
to the public Linux kernel mailing list in October 2022~\cite{ukllkml}. We had
several commenters with specific technical suggestions that can be readily
addressed and one maintainer with extensive and constructive feedback that we
will be incorporating. Only one maintainer seemed strongly opposed for
philosophical reasons.

To provide context for the size of the UKL patch, \cref{tab:patch} compares the
base UKL patch to a selection of Linux features described in Linux Weekly News
(LWN) \cite{lwn} articles in 2020. UKL's patch size is smaller, and it modifies
fewer files and Linux subsystems than many other patches accepted into Linux.
For comparison, the KML\cite{kml} patch, used in the recent Lupine work, which
runs applications in kernel mode, is 3177 LOC, a complexity that may have
contributed to the patch not being accepted upstream. While UKL goes beyond the
optimizations that KML does, both run applications in kernel mode and replace
\texttt{syscall}s with function calls. UKL base model goes beyond this and links
the applications with the kernel. So one question we had was why the
implementation of the UKL base model was so much simpler compared to KML. After
reviewing the code, we realized this simplicity is due to some fortuitous
changes since KML was introduced (discusses in \cref{sec:imp}). In addition, the
UKL base model supports only x64-64, while KML was introduced when it was
necessary to support i386 to be relevant.  Furthermore, the UKL base model does
not deal with older hardware, like the i8259 PIC, that had to be supported by
KML. The combined patch for the UKL base model and all the optimizations
explored in UKL so far is still only 1250 LOC (well below KML's). These
optimizations go beyond KML, enabling the broader specializations we discuss
later.

\begin{table}

\begin{tabular}{|l|r|r|r|l|}

\hline

\textbf{Project} & \textbf{LOC} & \textbf{Files} & \textbf{SubSys} &

\textbf{Outcome} \\

\hline

Popcorn  & 7763 & 64 & 14 & Out of tree \\

NetGPU  & 3827 & 45 & 14 & Rejected \\

DAMON  & 3805 & 24 & 3 & Accepted \\

\textbf{KML}   & 3177 & 70 & 16 & Out of tree \\

BPFStruct & 2639 & 32 & 10 & Accepted \\

BPFDump  & 2343 & 32 & 8 & Accepted \\

ArmMTE  & 1764 & 63 & 14 & Accepted \\

NFTOffload & 1579 & 56 & 24 & Accepted \\

\textbf{UKL} & 550 & 33 & 10 & - \\

KRSI   & 1085 & 29 & 11 & Accepted \\

LoopFS  & 891 & 27 & 5 & Rejected \\

FSGSBASE  & 562 & 16 & 9 & Accepted \\

BPFDisp  & 501 & 11 & 9 & Accepted \\

ArmAsym  & 370 & 13 & 9 & Rejected \\

BPFSleep  & 315 & 23 & 9 & Accepted \\

IOURestrictions & 194 & 2 & 2 & Accepted \\

CapPerfMon & 98 & 18 & 14 & Accepted \\

\hline

\end{tabular}

\caption{Comparison of the UKL base model patch to Kernel-Mode Linux (KML) and a
selection of Linux features described in Linux Weekly News (LWN) articles in
2020. We show patch size, files touched (how complex it is to reason about),
subsystems impacted (number of upstream kernel maintainers who need to review
and approve it), and the current status of the change.}

\label{tab:patch}

\vspace{-0.15in}

\end{table}

\subsection{Experimental Setup} \label{eval:setup} 

Unless otherwise stated, experiments are run on Dell R620 servers configured
with 128G of RAM across two sockets on a single NUMA node. Each socket contains
an Intel Xeon CPU E5-2660 0 @ 2.20GHz with 8 cores. The processors are
configured to disable Turbo Boost, hyper-threads, sleep states, and dynamic
frequency scaling. They are connected through a 10Gb link and use Broadcom
NetXtreme II BCM57800 1/10 Gigabit Ethernet NICs. Multi-node experiments use
identically configured nodes attached to the same top-of-rack switch to reduce
external noise. When comparing Linux and UKL, we use the same application,
kernel (5.14), and library versions for both (e.g., \texttt{glibc} version
2.31). Further, we use identical Linux configurations (modulo the UKL options)
and boot command line options. Both systems run with the following settings:
SMAP and SMEP disabled, and all Specter and Meltdown mitigations (e.g., KPTI)
disabled.

\subsection{Microbenchmarks} \label{eval:Micro}

Unikernels offer the opportunity to dramatically reduce the overhead of
interactions between the application and kernel code. We evaluate how UKL
optimizations impact the overhead of simple system calls (\cref{eval:sc-base}),
more expensive system calls (\cref{eval:sc-large}), and page faults
(\cref{eval:PFH}). Our results contradict recent work~\cite{lupine} that
suggests that the advantages are modest; we see that the reduction in overhead
is large (e.g., 90\%) and has a significant impact even for requests with large
payloads (e.g., 24\% with 8KB \texttt{recvfrom()}).

\subsubsection{System call base performance}\label{eval:sc-base} 

We measure the latency of five commonly used system calls, i.e.,
\texttt{getppid}, \texttt{read}, \texttt{write}, \texttt{sendto}, and
\texttt{recvfrom}. We compare the UKL base model and UKL\_BYP to unmodified
Linux. We use LEBench~\cite{lebench} to measure the latency of each system call
for each system at least 10,000 times. To ensure that we only measure the
intended operation, all buffers to store timing results are pre-allocated and
pre-faulted into memory. We do the same with buffers required for reading and
writing data for \texttt{read}, \texttt{write}, \texttt{sendto}, and
\texttt{recvfrom} system calls. 

\begin{figure}[]

\includegraphics[width=\columnwidth]{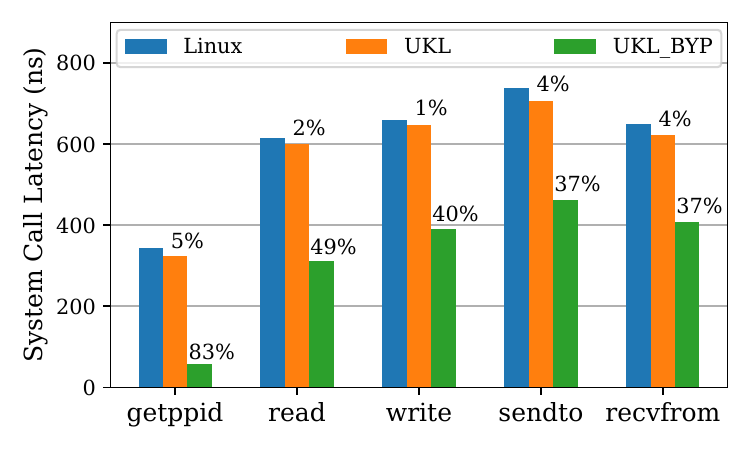}

\vspace{-0.15in}

\caption{Comparison of Linux, UKL base model, and UKL with bypass configuration
for simple system calls. With modern hardware, the UKL advantage of avoiding the
system call overhead is modest (<5\%). However, there appears to be a
significant advantage for simple calls with UKL\_BYP to avoid transition checks
between application and kernel code.} 

\label{fig:syscall}

\vspace{-0.15in}

\end{figure}

\Cref{fig:syscall} shows the results. As expected, the advantage of the
UKL base model over Linux is marginal (less than 5\%) because the
\texttt{syscall} instruction on modern systems is so optimized that replacing it
with \texttt{call} instruction does not provide huge benefits. The real win
comes with the UKL\_BYP configuration option, which shows, compared to Linux,
83\% improvement in \texttt{getppid}, 49\% for \texttt{read}, 40\% for
\texttt{write}, and 37\% for both \texttt{sendto} and \texttt{recvfrom} system
calls. These results show that Linux entry and exit code is the primary source
of latency in system calls, as opposed to the hardware cost of the
\texttt{syscall} instruction, and UKL\_BYP can greatly benefit workloads that
make many small system calls.


\begin{figure}[]

\includegraphics[width=\columnwidth]{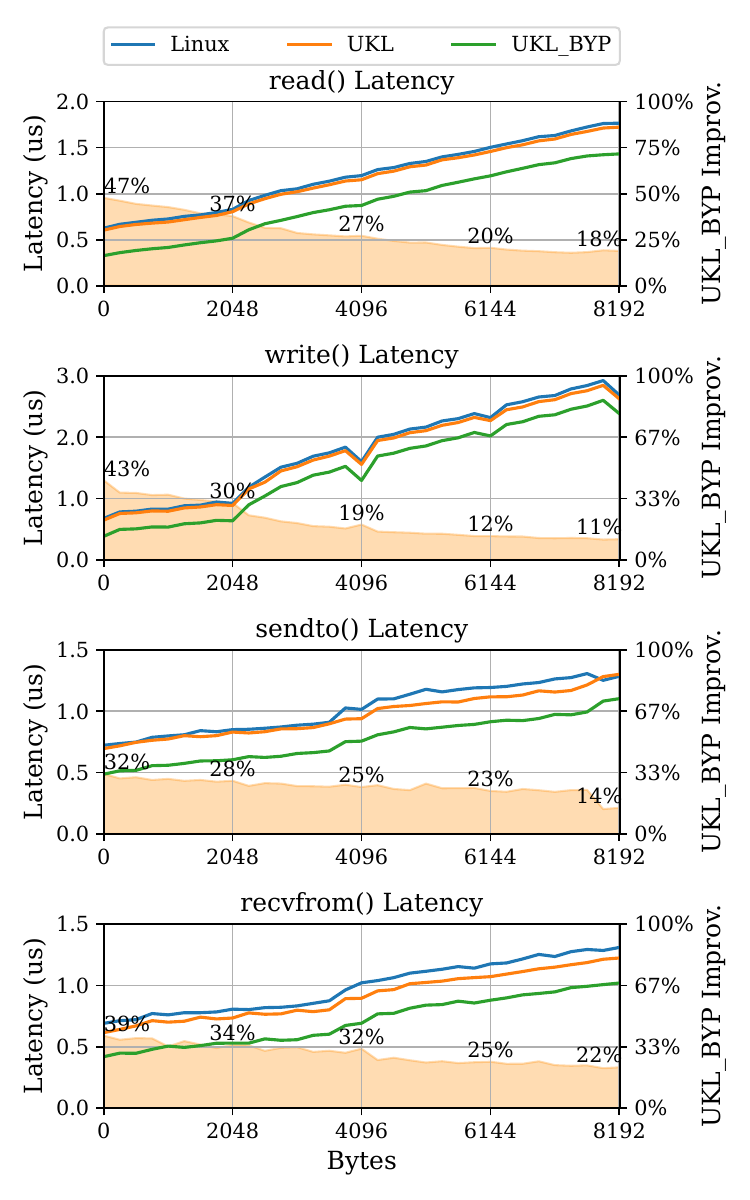}

\caption{Comparison of Linux, UKL base model, and UKL with bypass configuration
for four simple system calls. With increasing payload for each system call, UKL
shows modest improvement over Linux. But there is a significant advantage for
UKL, which bypasses the entry and exit code (UKL\_BYP). The orange area shows a
percentage improvement of UKL\_BYP over Linux, which decreases as the payload
increases but is still significant for the 8KB payload.}

\label{fig:foursyscalls} 

\vspace{-0.15in}

\end{figure}

\subsubsection{Large requests}\label{eval:sc-large} 

After evaluating base system call latency in \cref{eval:sc-base}, we now want to
measure the effect of increasing the input size on the latency of \texttt{read},
\texttt{write}, \texttt{sendto}, and \texttt{recvfrom} system calls. We again
use LEBench~\cite{lebench} to measure the latency. We increase the input size
from 1 byte to 8 KB, with 256-byte increments. The experiment is repeated
10,000 times for each size for each system (Linux, UKL base model, and
UKL\_BYP).

\Cref{fig:foursyscalls} shows that the UKL base model provides a negligible
performance improvement over Linux, and UKL\_BYP offers some performance
improvement (percentage improvement over Linux shown in the shaded region). The
performance improvement given by UKL\_BYP is due to bypassing the entry and exit
code and is not affected by the system call input size. We can see the
percentage improvement decreasing as the input size increases, but even for sizes
up to 8 KB, the percentage improvement is still significant, i.e., between
11\% and 22\%. This means that UKL\_BYP can also benefit workloads that make
system calls with larger payloads.

It is interesting to contrast our results with those from the recent Lupine
Linux \cite{lupine}, which shows (like us) that, other than a null system call,
the benefit of replacing \texttt{syscall} instruction with \texttt{call}
instruction is minimal (less than 5\%). From these results, authors of Lupine
Linux conclude that the benefit of optimizing the transition between application
and kernel code is minimal. But our results suggest that the major performance
gain comes not from eliminating the hardware cost but from eliminating all the
checks on the transition between the application and kernel code. Reducing this
overhead significantly impacts even expensive system calls.


\begin{figure*}[]

\includegraphics[width=\textwidth]{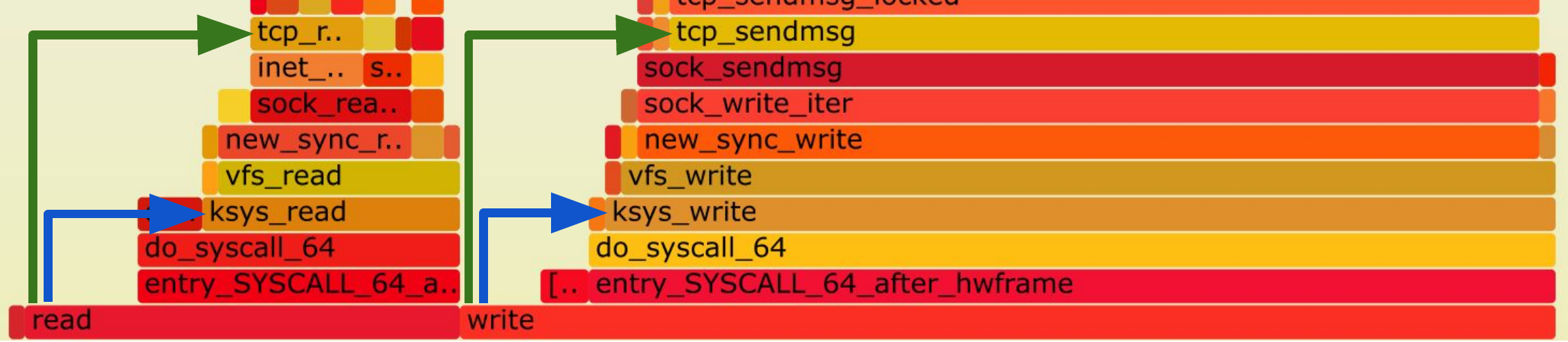}

\caption{Part of a flame graph generated after profiling Redis-UKL base model
with \texttt{perf}. The \texttt{read} and \texttt{write} functions at the bottom
reside in Redis code. Blue arrows show the code bypassed in UKL\_BYP, and green
arrows show deeper shortcuts. }

\label{fig:flamegraph} 

\vspace{-0.15in}

\end{figure*}

\subsubsection{\texttt{ret} versus \texttt{iret}}\label{eval:PFH}

\begin{table}[]

\begin{tabular}{|c|c|}

\hline

\textbf{\begin{tabular}[c]{@{}c@{}}No. of\\ Pages\end{tabular}} &
\textbf{\begin{tabular}[c]{@{}c@{}}UKL\_RET Improv.\\ over Linux\end{tabular}}
\\ \hline

1  & 7.66\% \\ \hline

2  & 9.14\% \\ \hline

3  & 8.34\% \\ \hline

4  & 9.65\% \\ \hline

5  & 9.85\% \\ \hline

6  & 11.65\%\\ \hline

7  & 10.82\%\\ \hline

8  & 11.88\%\\ \hline

9  & 11.30\%\\ \hline

10 & 11.45\%\\ \hline

\end{tabular}

\caption{Improvement of UKL\_RET configuration option, which uses \texttt{ret}
instruction to return from kernel to user code, over Linux, which uses the
\texttt{iret} instruction instead, with an increasing number of pages.}

\label{tab:ret-table}

\vspace{-0.25in}

\end{table}

In this section, we measure the improvement of the UKL\_RET
configuration option, which uses \texttt{ret} instruction to return from kernel
to user code, over Linux, which uses the \texttt{iret} instruction instead. We
used LEBench again to run a page fault benchmark. For both Linux and UKL\_RET, a
character array of size equal to one page was allocated, and time was measured
to write a character to the first index of that array, which would result in a
page fault. This would include a return from kernel to application code once the
page fault was handled. To increase the number of returns, the array was
incremented by a page size every time, and only the first index of every page
was written to, forcing a page fault on every write. The experiment was repeated
a thousand times for each array size. \Cref{tab:ret-table} shows the
percentage improvement of UKL\_RET over Linux, which shows the \texttt{ret}
instruction gives around 10\% improvement over the \texttt{iret} instruction.



\begin{table}[]

\centering

\begin{tabular}{|l|cc|cc|}

\hline

	& \multicolumn{2}{c|}{Read} & \multicolumn{2}{c|}{Write}  \\ \hline

System	& \multicolumn{1}{c|}{\begin{tabular}[c]{@{}c@{}}No.
of\\Ops\end{tabular}} & \begin{tabular}[c]{@{}c@{}}Tput \\ (Mb/s)\end{tabular} &
\multicolumn{1}{c|}{\begin{tabular}[c]{@{}c@{}}No. of\\Ops\end{tabular}} &
\begin{tabular}[c]{@{}c@{}}Tput \\ (Mb/s)\end{tabular} \\ \hline \hline

Linux	& \multicolumn{1}{c|}{324K} & 42.14 & \multicolumn{1}{c|}{323K} & 42.09
\\ \hline

UKL\_RET\_BYP & \multicolumn{1}{c|}{441K} & 57.37 & \multicolumn{1}{c|}{439K} &
57.2  \\ \hline

Improvement & \multicolumn{2}{c|}{36.1 \%} & \multicolumn{2}{c|}{35.9 \%}  \\
\hline

\end{tabular}

\caption{No. of operations completed and throughput in Mb/s of \texttt{fio} when
run with Linux and UKL\_RET\_BYP. UKL showed a 36\% improvement in operation
count and throughput.}

\label{tab:fio}

\vspace{-0.3in}

\end{table}

\subsection{I/O Latency}\label{eval:fio} 

Latency-sensitive applications, e.g., high-frequency trading, require high-speed
I/O. \Cref{eval:Micro} showed the impact of different optimizations on system
call latency. This section explores how those optimizations, and the resulting
speed-up in system call latency, impact I/O latency. 

To study I/O latency with UKL, we used the Flexible I/O
(\texttt{fio})~\cite{fio} benchmarking tool. We configured \texttt{fio} with an
I/O depth of 1 so that any speed-ups directly translate to latency gains,
performing randomly interleaved 4KB reads and writes using direct I/O to an 8GB
file for 30 seconds. This experiment was done on a 2021 Lenovo X1 laptop with
64G of RAM and a 1TB NVMe disk formatted with EXT4. Since each request has to
wait for the prior one to finish, improvement in the latency of the requests
directly translates into an increase in the number of requests serviced in a
fixed period. We compare the throughput of Linux with UKL\_RET\_BYP. As seen in
\cref{tab:fio}, UKL\_RET\_BYP provides a 36\% performance advantage over Linux
for read and write operations. This shows that UKL can have a large impact on
latency-sensitive applications.

\FloatBarrier

\subsection{Single Threaded Application - Redis}\label{eval:redis}

To understand the implication of UKL's design on applications, we evaluated it
with Redis, a widely used in-memory database. The simple design of Redis has
made it a popular target for unikernel research.

\Cref{fig:flamegraph} shows part of a flame graph \cite{10.1145/2909476}
we generated using \texttt{perf}. We see two clear opportunities for performance
improvement. Blue arrows show how we could shorten the execution path by
bypassing the entry and exit code for \texttt{read} and \texttt{write} system
calls and invoke the underlying functionality directly; that is, using UKL\_BYP.
The green arrows show that \texttt{read} and \texttt{write} calls, after all the
polymorphism, eventually translate into \texttt{tcp\_recvmsg} and
\texttt{tcp\_sendmsg} respectively. This suggests that further gains could be
obtained if we could create a shortcut that enabled an application like Redis
that always uses TCP to call the underlying routines directly.  

We developed a kernel call ``\texttt{shortcut}'' that would directly invoke the
underlying TCP routines. The change to Redis to invoke this routine required
only 10 LOC to be modified. 



\subsubsection{Comparison with other Systems} \label{redis-v}

In this section, we compare UKL's performance with Lupine~\cite{lupine} and
Unikraft~\cite{unikraft} using the experimental setup done by
Unikraft\cite{unikraft}; Redis server was deployed inside a single-core virtual
machine, and \texttt{redis-benchmark} deployed outside the virtual machine, on
the host, with 30 connections, 100k requests, and pipelining of 16 requests.
Since Lupine uses Linux 4.0 and the patch of UKL used for this evaluation is
based on Linux 5.14, we show those as baseline comparison points.

\Cref{tab:virtual-redis} shows that, compared to Linux 5.14, Unikraft performs
50\% better for \texttt{GET}s and 44\% better for \texttt{SET}s\footnote{We don't
see 70\% to 170\% improvement for Unikraft over Linux VM, as reported by authors
of Unikraft~\cite{unikraft}, because we turned off the security mitigations for
Linux.}. Linux 5.14 underperforms compared to Linux 4.0 due to differences in
compile-time configuration options, kernel versions 4.0 and 5.14, and in C
libraries (\texttt{musl} for Linux 4.0 and Lupine versus \texttt{glibc} for
Linux 5.14 and UKL). Lupine and the UKL base model show no advantage over their
respective baseline comparison points, showing the cost of a \texttt{syscall}
instruction is not high.

\begin{table}[]

\begin{tabular}{|l|c|c|c|}

\hline

\textbf{System}& \textbf{\begin{tabular}[c]{@{}c@{}}No. of\\ vCPUs\end{tabular}}
& \textbf{\begin{tabular}[c]{@{}c@{}}GET \\ (Mil. req/s)\end{tabular}} &
\textbf{\begin{tabular}[c]{@{}c@{}}SET\\ (Mil. req/s)\end{tabular}} \\ \hline
\hline

Unikraft & 1 & 1.08 & 0.91\\ \hline

\hline

Linux 4.0& 1 & 0.80 & 0.68\\ \hline

Lupine & 1 & 0.80 & 0.68\\ \hline

\hline

Linux 5.14 & 1 & 0.72 & 0.63\\ \hline

UKL base model & 1 & 0.71 & 0.62\\ \hline

UKL\_RET\_BYP& 1 & 0.72 & 0.63\\ \hline

\begin{tabular}[c]{@{}l@{}}UKL\_RET\_BYP\\ (shortcut)\end{tabular} & 1 & 0.77 &
0.66\\ \hline

\end{tabular}

\caption{Redis throughput comparison of UKL with Unikraft~\cite{unikraft} and
Lupine~\cite{lupine} on single-core VMs. Linux 4.0 is shown as a baseline
comparison for Lupine, and Linux 5.14 is shown as a baseline comparison for
UKL.}

\label{tab:virtual-redis}




\begin{tabular}{|l|c|c|c|}

\hline

\textbf{System}& \textbf{\begin{tabular}[c]{@{}c@{}}No. of\\ vCPUs\end{tabular}}
& \textbf{\begin{tabular}[c]{@{}c@{}}GET\\ (Mil. req/s)\end{tabular}} &
\textbf{\begin{tabular}[c]{@{}c@{}}SET\\ (Mil. req/s)\end{tabular}} \\ \hline
\hline

Linux 5.14 & 2 & 0.94& 0.77\\ \hline

\begin{tabular}[c]{@{}l@{}}UKL\_RET\_BYP\\ (shortcut)\end{tabular} & 2 & 1.00&
0.82\\ \hline

\end{tabular}

\caption{Repeat of the experiment in \cref{tab:virtual-redis}, except with Linux
5.14 and UKL\_RET\_BYP (shortcut) run in a 2-core virtual machine.}

\label{tab:virtual-redis-2core}

\vspace{-0.3in}

\end{table}

UKL\_RET\_BYP shows no improvement over Linux 5.14 and UKL\_RET\_BYP (shortcut)
has around 7\% improvement for \texttt{GET}s and around 5\% improvement for
\texttt{SET}s over Linux 5.14. With modest changes, UKL\_RET\_BYP (shortcut) is
only 29\% worse than Unikraft, a highly specialized unikernel written from
scratch, while preserving Linux's ecosystem and application and hardware
compatibility.

Using the \texttt{ps} utility on Linux and UKL showed many kernel background
threads that contend with Redis for CPU time. To remove this contention, we
re-ran the same experiment with two cores. Through Linux boot parameter
\texttt{isolcpus}, we isolated one of those cores so nothing would be scheduled
on that core, and using the \texttt{taskset} utility, we pinned Redis to that
core. \Cref{tab:virtual-redis-2core} show the results. Unikraft and Lupine are
not shown because they only support a single core. UKL\_RET\_BYP (shortcut)
gives more than 6\% better throughput for \texttt{SET}s and \texttt{GET}s
compared to Linux 5.14, and single core Unikraft is around 11\% better for
\texttt{SET}s and 8\% better for \texttt{GET}s. Adding a core is simple in UKL
since it preserves Linux's hardware compatibility, while adding that support can
be a huge engineering effort for a from-scratch unikernel.\footnote{It must be noted here
that all this performance advantage is not just due to removing CPU contention;
some might be due to interrupt processing parallelism, i.e., network interrupts
can now be serviced on two cores instead of one.}


\subsubsection{Bare Metal Experiment} \label{redis-bm}

While many unikernels can run Redis virtualized, few can run it bare-metal. In
this section, we evaluate the bare-metal performance advantages of Redis with
different configurations of UKL. For this experiment, instead of
\texttt{redis-benchmark}, used in \cref{redis-v}, we use the
\texttt{memtier\_benchmark} which generates a more realistic load than
\texttt{redis-benchmark}~\cite{memtier-bm} to fully drive Redis server.

\FloatBarrier

\begin{figure}[]

\includegraphics[width=\columnwidth]{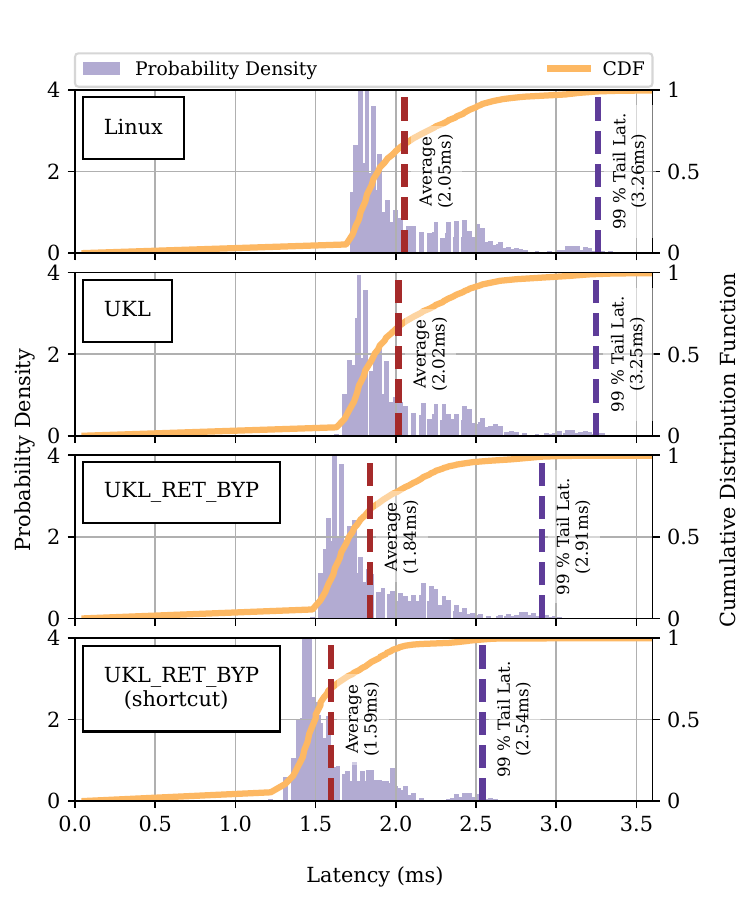}
\vspace{-0.3in}

\caption{Probability Density (purple bars) and CDF (orange line) of Redis
deployed on Linux, UKL, UKL\_RET\_BYP and UKL\_RET\_BYP (shortcut) and tested
with the \texttt{memtier\_benchmark}. Average latency (broken red line) and 99th
percentile tail latency (broken purple line) are also shown.}

\label{fig:redis-latency}

\vspace{-0.15in}

\end{figure}

We build and deploy unmodified Linux and different configurations of UKL on a
bare metal node to be the server (see \cref{eval:setup} for hardware and network
description) and connect it over a VLAN to another bare metal node running
Fedora 30, to host the \texttt{memtier\_benchmark} (client), which we configure
to create three threads, each creating 100 clients. Each client generates
100,000 requests for the Redis server. We also tried many other configurations
but had to find the one that would drive the Redis server as much as possible
without saturating the 10Gb network link between the client and the server.

\Cref{fig:redis-latency} shows the results of our experiment; we plot the
probability density and cumulative density function for Linux, UKL base model,
UKL\_RET\_BYP and UKL\_RET\_BYP (shortcut). Average and 99th percentile tail
latencies are also shown. The UKL base model shows a negligible advantage over
Linux. UKL\_RET\_BYP outperforms Linux and the UKL base model in average and
tail latency. \Cref{tab:redis-res} shows that UKL\_RET\_BYP has an 11\% better
tail latency and a 12\% improvement in throughput over Linux. Also,
UKL\_RET\_BYP (shortcut) outperforms both UKL\_RET\_BYP and Linux. UKL\_RET\_BYP
(shortcut) shows a 22\% improvement in tail latency and 26\% improvement in
throughput over Linux (\cref{tab:redis-res}). As with the virtual results, we
see that significant gains can be obtained if applications are modified to use custom
paths inside the kernel, e.g., UKL\_RET\_BYP (shortcut). Further, due to
virtualization overhead, UKL shows a larger performance improvement bare-metal
compared to virtualized deployment.

\begin{table}[]

\centering

\begin{tabular}{|l|cc|cc|}

\hline

\textbf{} & \multicolumn{2}{c|}{\textbf{99 \% tail lat}}  &
\multicolumn{2}{c|}{\textbf{Throughput}} \\ \hline

\textbf{System} & \multicolumn{1}{c|}{\textbf{(ms)}} & \textbf{Improv.} &
\multicolumn{1}{c|}{\textbf{(Kb/s)}} & \textbf{Improv.} \\ \hline \hline

Linux & \multicolumn{1}{c|}{3.26} & - & \multicolumn{1}{c|}{6375.20} & - \\
\hline

UKL base model & \multicolumn{1}{c|}{3.25} & 0.3\% &
\multicolumn{1}{c|}{6479.20} & 1.6\% \\ \hline

UKL\_RET\_BYP & \multicolumn{1}{c|}{2.91} & 11\% & \multicolumn{1}{c|}{7154.68}
& 12\% \\ \hline

\begin{tabular}[c]{@{}l@{}}UKL\_RET\_BYP\\ (shortcut)\end{tabular} &
\multicolumn{1}{c|}{2.54} & 22\% & \multicolumn{1}{c|}{8022.54} & 26\% \\ \hline

\end{tabular}

\caption{Redis throughput and latency improvements of UKL base model,
UKL\_RET\_BYP and UKL\_RET\_BYP (shortcut) over Linux}

\label{tab:redis-res}

\vspace{-0.3in}

\end{table}



\subsubsection{\texttt{perf} Analysis of Redis}

\begin{table}[]

\centering

\begin{tabularx}{\columnwidth}{|p{1.5cm}|*{4}{X|}}

\hline

\textbf{}& \textbf{Linux} & \textbf{UKL}&

\textbf{UKL\_RET\_BYP} & \textbf{UKL\_RET\_BYP (sc)} \\

\hline \hline

Instruc. & 358.61 B & 360.64 B & 357.84 B& 313.01 B \\ \hline

Reduc. vs Linux&& -0.57\%& 0.21\%& 12.72\% \\ \hline \hline

Cycles & 438.19 B & 440.12 B & 401.55 B& 342.39 B \\ \hline

Reduc. vs Linux&& -0.44\%& 8.36\%& 21.86\% \\ \hline \hline

Branches& 72.02 B & 72.37 B & 72.09 B & 63.16 B \\ \hline

Reduc. vs Linux & & -0.47\% & -0.09\% & 12.31\% \\ \hline

Branches Mispred. & 0.34 B& 0.42 B& 0.34 B& 0.09 B \\ \hline

\% Mispred. & 0.47\%& 0.58\%& 0.47\%& 0.15\% \\ \hline \hline

L1 dCache Accesses & 168.86 B & 170.69 B & 165.37 B& 141.77 B \\ \hline

Reduc. vs Linux && -1.08\%& 2.07\%& 16.04\% \\ \hline \hline

LLC Accesses& 6.07 B& 6.16 B& 5.54 B& 4.78 B \\ \hline

Reduc. vs Linux & & -1.44\% & 8.77\%& 21.28\% \\ \hline \hline

LLC Misses& 1.41 B& 1.4 B & 0.85 B& 0.79 B \\ \hline

Miss \% & 23.16\% & 22.72\% & 15.38\% & 16.59\% \\ \hline \hline

Time (s) & 200.87 & 201.85 & 184.04& 156.93 \\ \hline

Reduc. vs Linux&& -0.49\%& 8.38\%& 21.87\% \\ \hline \hline

Inst/Cycle & 0.82 & 0.82 & 0.89& 0.91 \\ \hline

Imp. over Linux&& 0.12\%& 8.89\% & 11.71\% \\ \hline

\end{tabularx}

\caption{Selected \texttt{perf} events for Redis}

\label{tab:redis-perf}

\vspace{-0.4in}

\end{table}

To better understand where the gains in Redis results (\cref{redis-bm}) come
from, we use \texttt{perf}~\cite{perf} to profile Redis on Linux and different
configurations of UKL. We re-run the same bare metal experiment described in
\cref{redis-bm}. The machine has a single NUMA node with two sockets, each
containing 8 cores. We disable hyperthreading, so each core runs a single
thread. Each core has a 32KB L1 instruction cache, a 32KB L1 data cache, and a
unified 256KB L2 cache. All 8 cores on a socket share a 20MB last-level cache.
We pin the Redis server on one of these cores through the \texttt{taskset}
utility. \Cref{tab:redis-perf} summarizes a few of the key results from
\texttt{perf}.

The UKL base model has a slightly higher number
of instructions than Linux. This is because it adds a few new instructions in
the transition code between the application and Linux kernel for environment
tracking and to ensure that the stored register state on the stack is correct
during those transitions. UKL\_RET\_BYP bypasses the entry and exit code which
gives it a slight improvement over Linux, and UKL\_RET\_BYP (shortcut) shows a
12.72 \% decrease in the number of instructions executed due to the deep
shortcuts. 

This difference in the number of instructions executed does not proportionally
translate into the number of cycles taken. UKL\_RET\_BYP and UKL\_RET\_BYP
(shortcut) show a much larger decrease in CPU cycles compared to the decrease in
the number of instructions executed. This means that bypassing those
instructions improved the efficiency of the remaining ones. Below we discuss
some potential reasons for that.

\Cref{tab:redis-perf} shows that the percentage decrease in the number of
branches encountered by each system is similar to the decrease in the number of
instructions executed. The percentage of mispredicted branches by UKL\_RET\_BYP
(shortcut), i.e., 0.15\%, is one-third of that of Linux, i.e., 0.47\%. 

\Cref{tab:redis-perf} shows that for L1 data cache accesses for UKL\_RET\_BYP
and UKL\_RET\_BYP (shortcut) show 2.07\% and 16.04\% decrease compared to Linux.
So not only did these systems execute fewer instructions (leading to a decrease
in L1 instruction cache accesses), but the fewer instructions also touched a
smaller percentage of data structures, etc., leading to a decrease in L1 data
cache accesses. A decrease in L1 instruction (due to fewer total instructions
executed) and data cache accesses would result in better cache efficiency for
the remaining contents of these caches, which would further result in improved
L2 performance as well, since it is a unified cache for both instructions and
data.

Cache efficiency is apparent in last-level cache (LLC) numbers in
\cref{tab:redis-perf}.  UKL\_RET\_BYP and UKL\_RET\_BYP (shortcut) show 8.77\%
and 21.28\% fewer LLC accesses compared to Linux. Remember that L1 and L2 are
per-core caches, but LLC is shared among 8 cores on a socket on this machine and
thus is expensive to access. The decrease in LLC accesses correlates with the
decrease in CPU cycles taken by UKL\_RET\_BYP  and UKL\_RET\_BYP (shortcut),
i.e., around 8\% and 21\%, respectively. UKL\_RET\_BYP and UKL\_RET\_BYP
(shortcut) also show fewer LLC misses compared to Linux, which further improves
their performance. But the main result is that not having to access LLC in the
first place due to better L1 and L2 cache efficiency might be a major factor in
UKL's performance.

The improvement in CPU cycles translates to an improvement in the time taken by
the Redis server to complete the requests, showing similar improvement, i.e.,
around 8\% for UKL\_RET\_BYP and around 21\% for UKL\_RET\_BYP (shortcut). A
fewer number of instructions executed and improved efficiency of the remaining
instructions lead to improved instructions per cycle, i.e., 8.9\% improvement
for UKL\_RET\_BYP and 11.7\% improvement for UKL\_RET\_BYP (shortcut).

This analysis helps us understand the existing performance improvements and
point to potential future directions that can result in the biggest
improvements. The fact that utilities like \texttt{perf} can run with UKL, but
not with other unikernels, is also an important result for UKL and shows how
preserving Linux's ecosystem helps developers, users, and operators bring the
shared knowledge over to UKL.



\subsection{Multi-threaded Application - Memcached}\label{eval:memcached} 

\begin{table}[]

\centering

\begin{tabular}{|*{4}{c|}}

\hline

\textbf{\begin{tabular}[c]{@{}c@{}}Conns. per\\Thread\end{tabular}}    &
\textbf{Linux} & \textbf{\begin{tabular}[c]{@{}c@{}}UKL\_RET\_BYP\\
(shortcut)\end{tabular}} & \textbf{ \% Improv.}\\ 

\hline \hline

1           & 0.22  & 0.22 & 0.36\%     \\ \hline

2           & 0.27  & 0.25 & 6.33\%     \\ \hline

3           & 0.33  & 0.29 & 10.76\%    \\ \hline

4           & 0.40  & 0.36 & 10.36\%    \\ \hline

5           & 0.50  & 0.47 & 7.30\%     \\ \hline

6           & 0.61  & 0.56 & 8.78\%     \\ \hline

7           & 0.71  & 0.64 & 9.90\%     \\ \hline

8           & 0.82  & 0.73 & 10.01\%    \\ \hline

9           & 0.94  & 0.84 & 10.60\%    \\ \hline

10          & 1.03  & 0.94 & 8.60\%     \\ \hline

\end{tabular}

\caption{99th percentile tail latency, in msec, of Memcached running on Linux
and UKL\_RET\_BYP (shortcut). Percentage improvement of
UKL\_RET\_BYP (shortcut) over Linux is also shown.
UKL\_RET\_BYP (shortcut) gets up to 10\% tail latency improvement over
Linux, even as the load on the Memcached server increases.}

\label{tab:memcached}

\vspace{-0.3in}

\end{table}

Many unikernels have shown results on Redis, a single-threaded application, but
only a few target a more complex application like Memcached~\cite{memcached}, a
multi-threaded key-value store that relies heavily on the pthread library and
\texttt{glibc}'s internal synchronization mechanisms and
libevent~\cite{libevent}, an event notification library that must be compiled
and linked with. 

We deploy Memcached like it is mostly deployed in data centers, i.e., in a VM,
and the client deployed on a separate physical node, sending requests over a
physical network. We deploy Memcached inside a 6-core VM; we ensure that all the
vCPUs are pinned to separate physical cores on the host. We configure Memcached
to run with 4 threads, each pinned to a separate vCPU. We compare Linux and
UKL\_RET\_BYP (shortcut), both built with \texttt{virtio} network
para-virtualization drivers. On the host end, we use the \texttt{vhost}
mechanism to share network queues between the host and the guest, taking the
QEMU userspace out of the critical path. This gives the VM a very high-speed
network. Inside the VM, we pin the \texttt{virtio} network queues to the
separate vCPUs as well to avoid cross-core interference of network interrupts.

We run the \texttt{memtier\_benchmark} on a separate physical node on Fedora 30.
We configure the benchmark to have 4 threads, each pinned to a separate physical
core and generating traffic of 100,000 requests. We increase the number of
connections per thread from 2 to 10, to increase the load on the Memcached
server and measure the 99th percentile tail latency. The 99th percentile tail
latency for Linux and UKL\_RET\_BYP (shortcut) is shown in
\cref{tab:memcached}, along with the percentage improvement over Linux, which
shows that UKL\_RET\_BYP (shortcut) has up to 10\% improvement over Linux, even
when experiencing higher load. Although UKL\_RET\_BYP (shortcut) had around 22\%
improvement in tail latency on Redis bare-metal, that number is now around 10\%.
This might be because Redis experiments measured latency improvements without
saturating the network, but this experiment is measuring latency while the
system is under heavy load and the network is also saturated. Also, this points
to the opportunity for further improvement for UKL in virtualized settings. 

\FloatBarrier


\section{Related Work} \label{sec:related}

There has been significant research on unikernels; we categorize them here as clean
slate designs, forks of existing operating systems, and incremental systems. 

\paragraph{Clean Slate Unikernels:} Many unikernel projects are written from
scratch or use a minimal kernel like MiniOS \cite{xen} for bootstrapping. These
systems have demonstrated improved security and smaller attack surfaces, e.g.,
Xax \cite{xax} and MirageOS \cite{mirageOS}, fast boot times, e.g., ClickOS
\cite{clickOS} and LightVM \cite{lightVM}, efficient memory use, e.g., OSv
\cite{osv}, and better runtime performance by application specialization, e.g.,
EbbRT \cite{ebbrt}, Unikraft \cite{unikraft},  SUESS \cite{seuss} and
Minicache\cite{minicache} The UKL effort was inspired by the tremendous results
demonstrated by clean slate unikernels. Our research targets trying to find ways
to integrate some of the advantages these systems have shown into a
general-purpose OS.

Some researchers have directly confronted the problem of compatibility, e.g.,
OSv \cite{osv} is almost Linux ABI compatible, and HermiTux is fully ABI
compatible with Linux binaries \cite{hermitux}. Other projects aim to make
building unikernels easier, e.g., EbbRT \cite{ebbrt}, Libra \cite{libra}, and
Unikraft \cite{unikraft}. We believe that UKL can adopt some of the ideas from
HermiTux to enable API compatibility and hope that some of the ideas of EbbRT
and Unikraft may point to a path to managing the complexity of evolving a
complex, customizable system. 

\paragraph{Forks of General Purpose Operating Systems:} Some projects either fork an existing
general-purpose OS code base or reuse a significant portion of one. Examples
include Drawbridge \cite{drawbridge}, which harvests code from Windows, Rump
kernel \cite{rumpkernel}, which uses NetBSD drivers, and Linux Kernel Library
(LKL) \cite{lkl}, which borrows code from Linux. Although constrained by the
original OS's design and structure, these systems generally have better
compatibility with existing applications \cite{drawbridge}. The codebases these
systems fork are well-tested \cite{lkl} and can serve as building blocks for
other research projects, e.g., Rump \cite{rumpkernel} has been used in other
projects \cite{librettos}. UKL builds on this research while attempting to find
a way to integrate unikernel optimizations in the OS/community. 

\paragraph{Incremental Systems:}  Kernel Mode Linux (KML) \cite{kml}, Lupine
\cite{lupine}, and X-Containers \cite{xcontainers} use an existing
general-purpose operating system (Linux) but, like UKL, try to maintain the
OS/community. UKL differs from these systems in going past reducing the hardware
cost of the transition between the application and kernel but also avoiding
software transition costs and enabling some of the deeper optimizations explored
by systems like EbbRT \cite{ebbrt} and Unikraft \cite{unikraft}. 

Lupine \cite{lupine} and X-Containers \cite{xcontainers} demonstrate
opportunities in customizing Linux through build time configurations, which is
orthogonal and complementary to UKL, and we hope to adapt some of their
optimizations in the future. 

Finally, UKL was motivated by our prior work~\cite{hotukl} that showed how a simple
TCP echo server could achieve major performance advantages when linked into the
Linux kernel. This work introduces the functionality needed to make it
possible to achieve those advantages for real applications on diverse systems,
including configurable features and optimizations, co-running userspace
processes, application preemption,  user library support, C and C++
constructors, execution mode tracking (user/kernel), stack switching, etc.


\section{Conclusion} \label{sec:conclusion} 

UKL demonstrates that a modern, general purpose, monolithic operating system
can be transformed into a system where unikernel optimizations become possible. 
Further, this can be done as a configuration option, in 550 lines of kernel changes.
The system can then be moved towards a highly specialized unikernel by applying 
different unikernel optimizations, some of which have been explored in this work.
In particular, we demonstrate how shortcutting can be used to allow
applications to interface directly with internal kernel paths. We show three
different system call entry points that trade off between specialization and performance.
We quantify optimization of real workloads, e.g., 26\% improvement in
Redis throughput while improving tail latency by 22\%.

UKL differs from existing unikernels. First, while application
and kernel code are statically linked together, UKL provides very different
execution environments for each, enabling applications to run in UKL with no
source modifications, and minimal changes to kernel invariants.
Second, UKL enables incremental performance optimization by modifying the application 
to take advantage of kernel capabilities directly, violating the traditional 
separation between kernel and application code. 
Third, user processes can run on top of UKL, enabling the
entire ecosystem of Linux tools and scripting to just work. Finally, it neither
follows a clean-slate approach, nor hard-forks an existing codebase.

We have only begun performance-optimizing UKL. A whole series of optimizations 
have become apparent beyond the current efforts, e.g., zero-copy interfaces, and 
cross layer link-time optimization.
From an application perspective, we believe that UKL will provide a natural path
for improving performance and reducing the complexity of concurrent
workloads. When user code moves into the kernel and runs at privilege, 
some operations might become faster, or possible in the first place, e.g.,
in a garbage collector, it might
be necessary to detect, or prevent concurrent accesses.
With easy and fast access to the memory infrastructure (e.g., page tables) and
the scheduler, many situations in which explicit, slow synchronization is needed
may be eliminated.

If the Linux community accepts UKL, we believe it will not only impact industrial
deployment, but will become a useful platform for future research. While the benefits of
broad application and hardware support may be obvious, a key contribution may be enabling
the reuse of existing testing and performance tools to study the inner workings of unikernels.


\section{Acknowledgments}

We want to thank the MOC~\cite{moc} industry partners, Red Hat and Two Sigma,
Han Dong, and the larger MOC and Red Hat engineering teams for their valuable
support, input and help with this research.   This research was made possible by
generous funding support from the Red Hat Collaboratory and NSF grant no.
1931714. We also thank our shepherd, Pierre Olivier, the anonymous reviewers and
program committee members of EuroSys'23, and our earlier submissions, whose
insightful comments and feedback improved the quality of our work.

\bibliographystyle{plain}
\bibliography{UKL.bib}

\begin{thebibliography}{10}

\bibitem{dpdk}
Dpdk - data plane development kit.
\newblock \url{https://www.dpdk.org/}.
\newblock Accessed on 2021-10-7.

\bibitem{libevent}
libevent – an event notification library.
\newblock \url{https://libevent.org/}.
\newblock (Accessed on 1/27/2023).

\bibitem{memtier-bm}
memtier\_benchmark: A high-throughput benchmarking tool for redis \& memcached.
\newblock
  \url{https://redis.com/blog/memtier_benchmark-a-high-throughput-benchmarking-tool-for-redis-memcached/}.
\newblock (Accessed on 2/15/2023).

\bibitem{moc}
The moc alliance.
\newblock \url{https://massopen.cloud/}.
\newblock (Accessed on 3/13/2023).

\bibitem{perf}
perf: Linux profiling with performance counters.
\newblock \url{https://perf.wiki.kernel.org/index.php/Main_Page}.
\newblock (Accessed on 1/27/2023).

\bibitem{ukllkml}
[rfc ukl 00/10] unikernel linux (ukl).
\newblock
  \url{https://lore.kernel.org/lkml/20221003222133.20948-1-aliraza@bu.edu/}.
\newblock (Accessed on 12/28/2022).

\bibitem{spdk}
{Storage Performance Development Kit}.
\newblock \url{https://spdk.io/}, 2018.
\newblock (Accessed on 01/16/2019).

\bibitem{nitro}
{Amazon}.
\newblock \url{https://aws.amazon.com/ec2/nitro/}, 2022.
\newblock (Accessed 10/19/2022).

\bibitem{libra}
Glenn Ammons, Jonathan Appavoo, Maria Butrico, Dilma Da~Silva, David Grove,
  Kiyokuni Kawachiya, Orran Krieger, Bryan Rosenburg, Eric Van~Hensbergen, and
  Robert~W Wisniewski.
\newblock Libra: a library operating system for a jvm in a virtualized
  execution environment.
\newblock In {\em Proceedings of the 3rd international conference on Virtual
  execution environments}, pages 44--54, 2007.

\bibitem{andersoncase}
Thomas~E Anderson.
\newblock {\em The case for application-specific operating systems}.
\newblock University of California, Berkeley, Computer Science Division, 1993.

\bibitem{fio}
Jens Axboe.
\newblock \url{https://fio.readthedocs.io/en/latest/fio_doc.html}.
\newblock (Accessed on 10/13/2022).

\bibitem{xen}
Paul Barham, Boris Dragovic, Keir Fraser, Steven Hand, Tim Harris, Alex Ho,
  Rolf Neugebauer, Ian Pratt, and Andrew Warfield.
\newblock Xen and the art of virtualization.
\newblock {\em ACM SIGOPS operating systems review}, 37(5):164--177, 2003.

\bibitem{DBLP:journals/corr/BeamerAP15}
Scott Beamer, Krste Asanovic, and David~A. Patterson.
\newblock The {GAP} benchmark suite.
\newblock {\em CoRR}, abs/1508.03619, 2015.

\bibitem{seuss}
James Cadden, Thomas Unger, Yara Awad, Han Dong, Orran Krieger, and Jonathan
  Appavoo.
\newblock Seuss: skip redundant paths to make serverless fast.
\newblock In {\em Proceedings of the Fifteenth European Conference on Computer
  Systems}, pages 1--15, 2020.

\bibitem{cachekernel}
David~R Cheriton and Kenneth~J Duda.
\newblock A caching model of operating system kernel functionality.
\newblock {\em ACM SIGOPS Operating Systems Review}, 29(1):83--86, 1995.

\bibitem{xax}
John~R Douceur, Jeremy Elson, Jon Howell, and Jacob~R Lorch.
\newblock Leveraging legacy code to deploy desktop applications on the web.
\newblock In {\em OSDI}, volume~8, pages 339--354, 2008.

\bibitem{exokernel}
Dawson~R Engler, M~Frans Kaashoek, and James O'Toole~Jr.
\newblock Exokernel: An operating system architecture for application-level
  resource management.
\newblock {\em ACM SIGOPS Operating Systems Review}, 29(5):251--266, 1995.

\bibitem{10.1145/2909476}
Brendan Gregg.
\newblock The flame graph.
\newblock {\em Commun. ACM}, 59(6):48–57, may 2016.

\bibitem{intel}
{Intel}.
\newblock \url{https://www.dpdk.org/}, 2010.
\newblock (Accessed 01/17/2019).

\bibitem{rumprun}
Antti Kantee.
\newblock {\em The design and implementation of the anykernel and rump
  kernels}.
\newblock 2nd edition, 2016.

\bibitem{rumpkernel}
Antti Kantee et~al.
\newblock Flexible operating system internals: the design and implementation of
  the anykernel and rump kernels.
\newblock 2012.

\bibitem{osv}
Avi Kivity, Dor Laor, Glauber Costa, Pekka Enberg, Nadav Har’El, Don Marti,
  and Vlad Zolotarov.
\newblock Osv—optimizing the operating system for virtual machines.
\newblock In {\em 2014 $\{$USENIX$\}$ Annual Technical Conference
  ($\{$USENIX$\}$$\{$ATC$\}$ 14)}, pages 61--72, 2014.

\bibitem{serverlessenddominance}
Ricardo Koller and Dan Williams.
\newblock {Will Serverless End the Dominance of Linux in the Cloud?}
\newblock In {\em Proceedings of the 16th Workshop on Hot Topics in Operating
  Systems}, pages 169--173. ACM, 2017.

\bibitem{unikraft}
Simon Kuenzer, Vlad-Andrei B{\u{a}}doiu, Hugo Lefeuvre, Sharan Santhanam,
  Alexander Jung, Gaulthier Gain, Cyril Soldani, Costin Lupu, {\c{S}}tefan
  Teodorescu, Costi R{\u{a}}ducanu, et~al.
\newblock Unikraft: fast, specialized unikernels the easy way.
\newblock In {\em Proceedings of the Sixteenth European Conference on Computer
  Systems}, pages 376--394, 2021.

\bibitem{minicache}
Simon Kuenzer, Anton Ivanov, Filipe Manco, Jose Mendes, Yuri Volchkov, Florian
  Schmidt, Kenichi Yasukata, Michio Honda, and Felipe Huici.
\newblock Unikernels everywhere: The case for elastic cdns.
\newblock In {\em Proceedings of the 13th ACM SIGPLAN/SIGOPS International
  Conference on Virtual Execution Environments}, pages 15--29, 2017.

\bibitem{lupine}
Hsuan-Chi Kuo, Dan Williams, Ricardo Koller, and Sibin Mohan.
\newblock A linux in unikernel clothing.
\newblock In {\em Proceedings of the Fifteenth European Conference on Computer
  Systems}, pages 1--15, 2020.

\bibitem{nemesis}
Ian~M. Leslie, Derek McAuley, Richard Black, Timothy Roscoe, Paul Barham, David
  Evers, Robin Fairbairns, and Eoin Hyden.
\newblock The design and implementation of an operating system to support
  distributed multimedia applications.
\newblock {\em IEEE journal on selected areas in communications},
  14(7):1280--1297, 1996.

\bibitem{secrecy}
John Liagouris, Vasiliki Kalavri, Muhammad Faisal, and Mayank Varia.
\newblock Secrecy: Secure collaborative analytics in untrusted clouds.
\newblock {\em to apear NSDI 2023}, 2023.

\bibitem{mirageOS}
Anil Madhavapeddy, Richard Mortier, Charalampos Rotsos, David Scott, Balraj
  Singh, Thomas Gazagnaire, Steven Smith, Steven Hand, and Jon Crowcroft.
\newblock Unikernels: Library operating systems for the cloud.
\newblock {\em ACM SIGARCH Computer Architecture News}, 41(1):461--472, 2013.

\bibitem{kml}
Toshiyuki Maeda and Akinori Yonezawa.
\newblock Kernel mode linux: Toward an operating system protected by a type
  theory.
\newblock In {\em Annual Asian Computing Science Conference}, pages 3--17.
  Springer, 2003.

\bibitem{lightVM}
Filipe Manco, Costin Lupu, Florian Schmidt, Jose Mendes, Simon Kuenzer, Sumit
  Sati, Kenichi Yasukata, Costin Raiciu, and Felipe Huici.
\newblock My vm is lighter (and safer) than your container.
\newblock In {\em Proceedings of the 26th Symposium on Operating Systems
  Principles}, pages 218--233, 2017.

\bibitem{clickOS}
Joao Martins, Mohamed Ahmed, Costin Raiciu, Vladimir Olteanu, Michio Honda,
  Roberto Bifulco, and Felipe Huici.
\newblock Clickos and the art of network function virtualization.
\newblock In {\em 11th $\{$USENIX$\}$ symposium on networked systems design and
  implementation ($\{$NSDI$\}$ 14)}, pages 459--473, 2014.

\bibitem{rcu}
Paul~E McKenney and John~D Slingwine.
\newblock Read-copy update: Using execution history to solve concurrency
  problems.
\newblock In {\em Parallel and Distributed Computing and Systems}, volume
  509518, 1998.

\bibitem{memcached}
Memcached.
\newblock \url{https://memcached.org/}.
\newblock (Accessed on 05/30/2022).

\bibitem{bluegene}
Jos{\'e} Moreira, Michael Brutman, Jos{\'e} Castanos, Thomas Engelsiepen, Mark
  Giampapa, Tom Gooding, Roger Haskin, Todd Inglett, Derek Lieber, Pat
  McCarthy, et~al.
\newblock Designing a highly-scalable operating system: The blue gene/l story.
\newblock In {\em Proceedings of the 2006 ACM/IEEE conference on
  Supercomputing}, pages 118--es, 2006.

\bibitem{bolted}
Amin Mosayyebzadeh, Apoorve Mohan, Sahil Tikale, Mania Abdi, Nabil Schear,
  Trammell Hudson, Charles Munson, Larry Rudolph, Gene Cooperman, Peter
  Desnoyers, and Orran Krieger.
\newblock Supporting security sensitive tenants in a {Bare-Metal} cloud.
\newblock In {\em 2019 USENIX Annual Technical Conference (USENIX ATC 19)},
  pages 587--602, Renton, WA, July 2019. USENIX Association.

\bibitem{lwn}
Linux~Weekly News.
\newblock \url{https://lwn.net/}.
\newblock (Accessed on 05/30/2022).

\bibitem{nginx}
Nginx.
\newblock \url{https://nginx.org/}.
\newblock (Accessed on 10/13/2022).

\bibitem{librettos}
Ruslan Nikolaev, Mincheol Sung, and Binoy Ravindran.
\newblock Librettos: a dynamically adaptable multiserver-library os.
\newblock In {\em Proceedings of the 16th ACM SIGPLAN/SIGOPS International
  Conference on Virtual Execution Environments}, pages 114--128, 2020.

\bibitem{hermitux}
Pierre Olivier, Daniel Chiba, Stefan Lankes, Changwoo Min, and Binoy Ravindran.
\newblock A binary-compatible unikernel.
\newblock In {\em Proceedings of the 15th ACM SIGPLAN/SIGOPS International
  Conference on Virtual Execution Environments}, pages 59--73, 2019.

\bibitem{drawbridge}
Donald~E Porter, Silas Boyd-Wickizer, Jon Howell, Reuben Olinsky, and Galen~C
  Hunt.
\newblock Rethinking the library os from the top down.
\newblock In {\em Proceedings of the sixteenth international conference on
  Architectural support for programming languages and operating systems}, pages
  291--304, 2011.

\bibitem{lkl}
Octavian Purdila, Lucian~Adrian Grijincu, and Nicolae Tapus.
\newblock Lkl: The linux kernel library.
\newblock In {\em 9th RoEduNet IEEE International Conference}, pages 328--333.
  IEEE, 2010.

\bibitem{hotukl}
Ali Raza, Parul Sohal, James Cadden, Jonathan Appavoo, Ulrich Drepper, Richard
  Jones, Orran Krieger, Renato Mancuso, and Larry Woodman.
\newblock Unikernels: The next stage of linux's dominance.
\newblock In {\em Proceedings of the Workshop on Hot Topics in Operating
  Systems}, HotOS '19, page 7–13, New York, NY, USA, 2019. Association for
  Computing Machinery.

\bibitem{redis}
Redis.
\newblock \url{https://redis.io/}.
\newblock (Accessed on 05/30/2022).

\bibitem{lebench}
Xiang Ren, Kirk Rodrigues, Luyuan Chen, Camilo Vega, Michael Stumm, and Ding
  Yuan.
\newblock An analysis of performance evolution of linux's core operations.
\newblock In {\em Proceedings of the 27th ACM Symposium on Operating Systems
  Principles}, pages 554--569, 2019.

\bibitem{netmap}
Luigi Rizzo.
\newblock netmap: a novel framework for fast packet i/o.
\newblock In {\em 21st USENIX Security Symposium (USENIX Security 12)}, pages
  101--112, 2012.

\bibitem{ebbrt}
Dan Schatzberg, James Cadden, Han Dong, Orran Krieger, and Jonathan Appavoo.
\newblock Ebbrt: A framework for building per-application library operating
  systems.
\newblock In {\em 12th $\{$USENIX$\}$ Symposium on Operating Systems Design and
  Implementation ($\{$OSDI$\}$ 16)}, pages 671--688, 2016.

\bibitem{xcontainers}
Zhiming Shen, Zhen Sun, Gur-Eyal Sela, Eugene Bagdasaryan, Christina
  Delimitrou, Robbert Van~Renesse, and Hakim Weatherspoon.
\newblock X-containers: Breaking down barriers to improve performance and
  isolation of cloud-native containers.
\newblock In {\em Proceedings of the Twenty-Fourth International Conference on
  Architectural Support for Programming Languages and Operating Systems}, pages
  121--135, 2019.

\bibitem{serverlesspeeking}
Liang Wang, Mengyuan Li, Yinqian Zhang, Thomas Ristenpart, and Michael Swift.
\newblock {Peeking Behind the Curtains of Serverless Platforms}.
\newblock In {\em Proceedings of the 2018 USENIX Conference on Usenix Annual
  Technical Conference}, USENIX ATC '18, pages 133--145, Berkeley, CA, USA,
  2018. USENIX Association.

\bibitem{nabla}
Dan Williams, Ricardo Koller, Martin Lucina, and Nikhil Prakash.
\newblock Unikernels as processes.
\newblock In {\em Proceedings of the ACM Symposium on Cloud Computing}, pages
  199--211, 2018.

\end{thebibliography}

\end{document}